\input{aipcheck}

\documentclass[
    ,final            % use final for the camera ready runs
  ]
  {aipproc}

\layoutstyle{8x11single}

\usepackage{amsmath,amssymb}
\usepackage{graphicx}
\usepackage{slashed}
\usepackage[T1]{fontenc}
\usepackage{color}
\usepackage{times}
\usepackage{overpic}
\usepackage{rotating}
\usepackage{subfig}

\begin{document}

\title{Functional renormalization group study \\ of nuclear and neutron matter}


\author{Matthias Drews}{
  address={Physik Department, Technische Universit\"{a}t M\"{u}nchen, D-85747 Garching, Germany},
  altaddress={ECT*, Villa Tambosi, I-38123 Villazzano (Trento), Italy}
}

\author{Wolfram Weise}{
  address={Physik Department, Technische Universit\"{a}t M\"{u}nchen, D-85747 Garching, Germany},
  altaddress={ECT*, Villa Tambosi, I-38123 Villazzano (Trento), Italy}
}

\begin{abstract}
	A chiral model based on nucleons interacting via boson exchange is investigated. Fluctuation effects are included consistently beyond the mean-field approximation in the framework of the functional renormalization group. The liquid-gas phase transition of symmetric nuclear matter is studied in detail. No sign of a chiral restoration transition is found up to temperatures of about 100\,MeV and densities of at least three times the density of normal nuclear matter. Moreover, the model is extended to asymmetric nuclear matter and the constraints from neutron star observations are discussed.
\end{abstract}

\maketitle

%%%%%%%%%%%%%%%%%%%%%%%%%%%%%%%%%%%%%%%%%%%%
%% MAINMATTER
%%%%%%%%%%%%%%%%%%%%%%%%%%%%%%%%%%%%%%%%%%%%

\section{Introduction}

The phase diagram of Quantum Chromodynamics (QCD) includes various facets from the early universe to heavy-ion collisions and neutron stars. One important aspect of QCD in its hadronic phase is spontaneous chiral symmetry breaking. Whereas a crossover to a chirally restored phase at zero chemical potential is well established, the situation at higher chemical potentials is less clear. A first-order phase transition with a critical endpoint has been intensively discussed in the literature. Model calculations often predict a chiral first-order transition close to the liquid-gas phase transition at relatively low baryon densities. We address this problem by studying a chiral nucleon-meson (ChNM) model \cite{berges2000chiral,berges2003quark} with parameters fitted to the empirical properties of nuclear matter close to the liquid-gas transition. In addition to an earlier mean-field analysis \cite{floerchinger2012chemical}, we include field fluctuations beyond the mean-field approximations in the framework of the functional renormalization group (FRG). An important finding of our study of the FRG-improved chiral nucleon-meson (FRG-ChNM) model is that chiral symmetry restoration in symmetric nuclear matter is shifted up to very high densities.

A similar issue shows up also in the discussion of pure neutron matter. Perturbative in-medium chiral effective field theory (ChEFT) calculations show an almost linearly decreasing chiral condensate as a function of density \cite{kaiser2009chiral,krueger2013neutron}. Taken at face value, chiral symmetry would be restored already at about three times nuclear saturation density. This would imply a possibly ``exotic'' composition of matter in the central core of heavy neutron stars at characteristic densities around and beyond five times nuclear saturation density. In contrast, the FRG-ChNM model predicts that the chiral order parameter is stabilized by fluctuations up to densities of about seven times nuclear saturation density.
Consequently, chiral symmetry is still spontaneously broken at high densities and the model is applicable in order to provide an equation of state as input for neutron star studies. We find that the mass and radius constraints from recent neutron star observations are fulfilled.

\section{Chiral nucleon-meson model}
The ChNM model \cite{berges2000chiral,berges2003quark} is based on a linear sigma model and involves an effective potential that is adjusted to reproduce empirical properties of nuclear matter around the liquid-gas phase transition. The relevant degrees of freedom in this regime are nucleons, i.e., protons and neutrons, combined in a two-component field $\psi=(\psi_p,\psi_n)$. The long-range nucleon-nucleon interaction is generated by pion exchange. The isovector pion is combined with a scalar-isoscalar field, $\sigma$, in a chiral four-component field $(\vec\pi, \sigma)$. 
Short-range dynamics is modeled by isoscalar-vector interactions, \mbox{$(\bar\psi\gamma_\mu\psi)\,(\bar\psi\gamma^\mu\psi)$}, and isovector-vector interactions, \mbox{$(\bar\psi\gamma_\mu\vec\tau\psi)\cdot(\bar\psi\gamma^\mu\vec\tau\psi)$}, where $\vec\tau$ are the isospin Pauli matrices. After a Hubbard--Stratonovich transformation, these interactions can be thought of as mediated by exchange of heavy vector bosons, $\omega_\mu$ and $\vec\rho_\mu$, respectively. The corresponding field strength tensors are \mbox{$F_{\mu\nu}^{(\omega)}=\partial_\mu\omega_\nu-\partial_\nu\omega_\mu$} and \mbox{$\vec F_{\mu\nu}^{(\rho)}=\partial_\mu\vec\rho_\nu-\partial_\nu\vec\rho_\mu-g_\rho\,\vec\rho_\mu\times\vec\rho_\nu$}. The vector bosons are heavy as compared to the relevant scales and will be treated as background fields in a mean-field approximation. Rotational invariance implies that only the time components $\omega_0$ and $\rho_0^3$ are non-vanishing, and the non-abelian part of $\vec F_{\mu\nu}^{(\rho)}$ vanishes. In Minkowski space, the Lagrangian of the ChNM model is given by
\begin{align}
	\begin{aligned}
		\mathcal L&=\bar\psi i\gamma_\mu\partial^\mu\psi+\frac 12\partial_\mu\sigma\,\partial^\mu\sigma+\frac 12\partial_\mu\vec\pi\cdot\partial^\mu\vec\pi -\bar\psi\Big[g(\sigma+i\gamma_5\,\vec\tau\cdot\vec\pi)+\gamma_\mu(g_\omega\, \omega^\mu+g_\rho\vec\tau\cdot\vec\rho^\mu)\Big]\psi \\
		&\quad-\frac 14 F^{(\omega)}_{\mu\nu}F^{(\omega)\mu\nu} - \frac 14 \vec F^{(\rho)}_{\mu\nu}\cdot\vec F^{(\rho)\mu\nu} +\frac 12m_v^2\big(\omega_\mu\,\omega^\mu+\,\vec\rho_\mu\cdot\vec\rho^\mu\big)- {\cal U}(\sigma,\vec\pi).
	\end{aligned}
\end{align}
In a mean-field approximation, the fermions are integrated out and the bosons are replaced by their non-vanishing vacuum expectation values, denoted for convenience by $\sigma$, $\omega_0$ and $\rho_0^3$. The expectation value of $\sigma$ generates a dynamical nucleon mass, $M_N=g\sigma$, and the expectation values of the vector fields shift the neutron and proton chemical potentials according to
\begin{align}
	\mu_{n,p}^{\text{eff}}=\mu_{n,p}-g_\omega\omega_0\pm g_\rho \rho_0^3\,.
\end{align}
The parameters of the potential ${\cal U}$ together with the vector boson coupling strengths are fitted to reproduce empirical data, such as the binding energy, the compression modulus and the surface tension of nuclear matter, the symmetry energy, as well as the nucleon mass, the pion mass and the pion decay constant \cite{floerchinger2012chemical,drews2013thermodynamic,drews2014functional}.

Fluctuations are included in the non-perturbative framework of the functional renormalization group. The flow of an effective action $\Gamma_k$ (which depends on a renormalization scale $k$) is determined in such a way that it interpolates between the ultraviolet action $\Gamma_{k=\Lambda}=S$ at a cutoff $\Lambda = 1.4\text{\,GeV}$ and the full quantum effective action $\Gamma_{k=0}=\Gamma_{\text{eff}}$. The respective flow equation \cite{wetterich1993exact} is given by:
\begin{align}
	\begin{aligned}
		k\,\frac{\partial\Gamma_k}{\partial k}=
		\begin{aligned}
			\vspace{1cm}
			\includegraphics[width=0.08\textwidth]{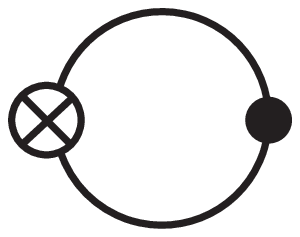}
		\end{aligned} \vspace{-1cm}=\frac 12 \operatorname{Tr} \bigg[k\,\frac{\partial R_k}{\partial k}\cdot\Big(\Gamma_k^{(2)}+R_k\Big)^{-1}\bigg]\,.
	\end{aligned}
\end{align}
In this formula, $\Gamma_k^{(2)}$ is the second derivative of the effective action with respect to the fields. Moreover, $R_k$ is a regulator function that makes the flow IR finite. Note that $\Gamma_k^{(2)}+R_k$ is the full inverse propagator. The flow equation is, in principle, an exact equation that gives the full non-perturbative result for the quantum effective action.

As demonstrated in Refs.~\cite{drews2013thermodynamic,drews2014dense,drews2014functional}, the fluctuations of nucleons, pions and the $\sigma$ field are included with respect to the potential at the liquid-gas phase transition at vanishing temperature. The potential in the ultraviolet is chosen such that the (slightly modified) mean-field potential is reproduced at that point. The loops include pionic fluctuations and important particle-hole excitations of the nucleons around the Fermi surface. The flow equation sums up in a non-perturbative way all kinds of pion exchange processes as well as multi-nucleon forces that are generated by the effective potential 
${\cal U}(\sigma)$ through its highly nonlinear dependence on the expectation value of the $\sigma$ field.

By varying the temperature and the proton and neutron chemical potentials, the nuclear phase diagram around the liquid-gas transition can then be studied as demonstrated in Refs.~\cite{drews2013thermodynamic,drews2014functional}.

\section{Results and discussion}
\begin{figure}
	\centering
	\begin{overpic}[width=0.4\textwidth]{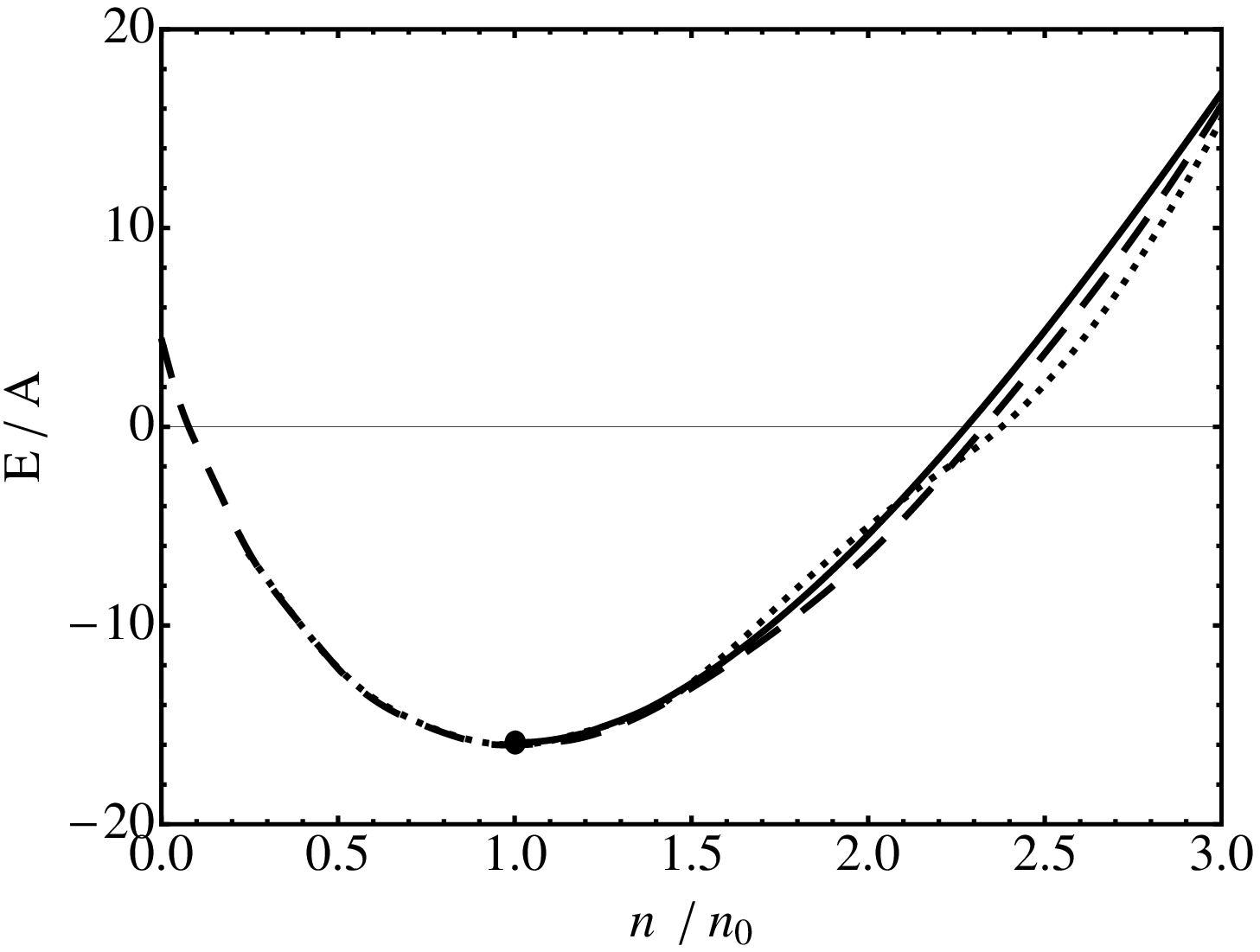}
		\put(3,49){\begin{rotate}{90} (MeV) \end{rotate}}
	\end{overpic}
	\qquad
	\begin{overpic}[width=0.4\textwidth]{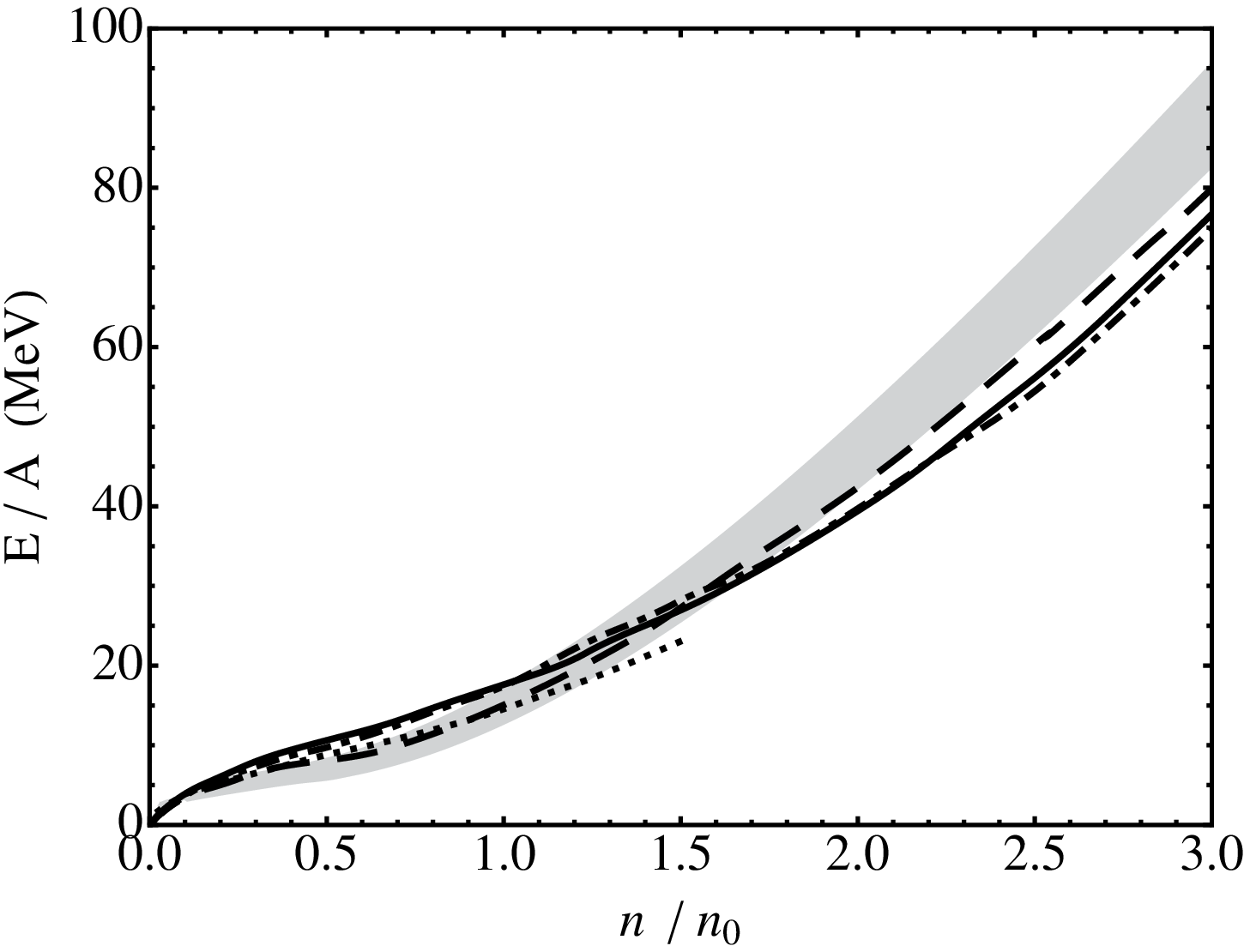}
	\end{overpic}
%	\vspace{-0.2cm}
	\caption{The energy per particle as a function of density (normalized to nuclear saturation density $n_0$). Left: symmetric nuclear matter, with the FRG-ChNM result (solid line), the APR EoS (dotted, \cite{akmal1998equation}) and a QMC computation (dashed, \cite{armani2011recent}). Right: pure neutron matter, with the FRG-ChNM result (gray band, with $29\text{\,MeV}\le E_{\text{sym}} \le 33\text{\,MeV}$), ChEFT (full line, \cite{hell2014dense}), QMC based on realistic potentials (dashed, \cite{armani2011recent}), QMC based on chiral potentials (dotted, \cite{roggero2014quantum}), and APR (dashed-dotted, \cite{akmal1998equation}).}
	\label{fig:EoS}
\end{figure}

Figure\,\ref{fig:EoS} shows the energy per particle of nuclear matter and pure neutron matter at vanishing temperature. The symmetry energy is not very accurately determined \cite{lattimer2013constraining}, so we have varied the parameters in order to cover a range $29\text{\,MeV}\le E_{\text{sym}} \le 33\text{\,MeV}$ represented by the band for pure neutron matter. Both the equation of state of symmetric nuclear matter and of pure neutron matter are close to results of sophisticated many-body calculations. The fluctuations beyond mean-field approximation play an important role in improving the agreement, as compared to mean-field results. As a consequence, the FRG-ChNM model is well prepared to cover a broad range of densities up to at least three times nuclear saturation density, $n_0=0.16\text{\,fm}^{-3}$.

\begin{figure}
	\centering
		\begin{overpic}[width=0.4\textwidth]{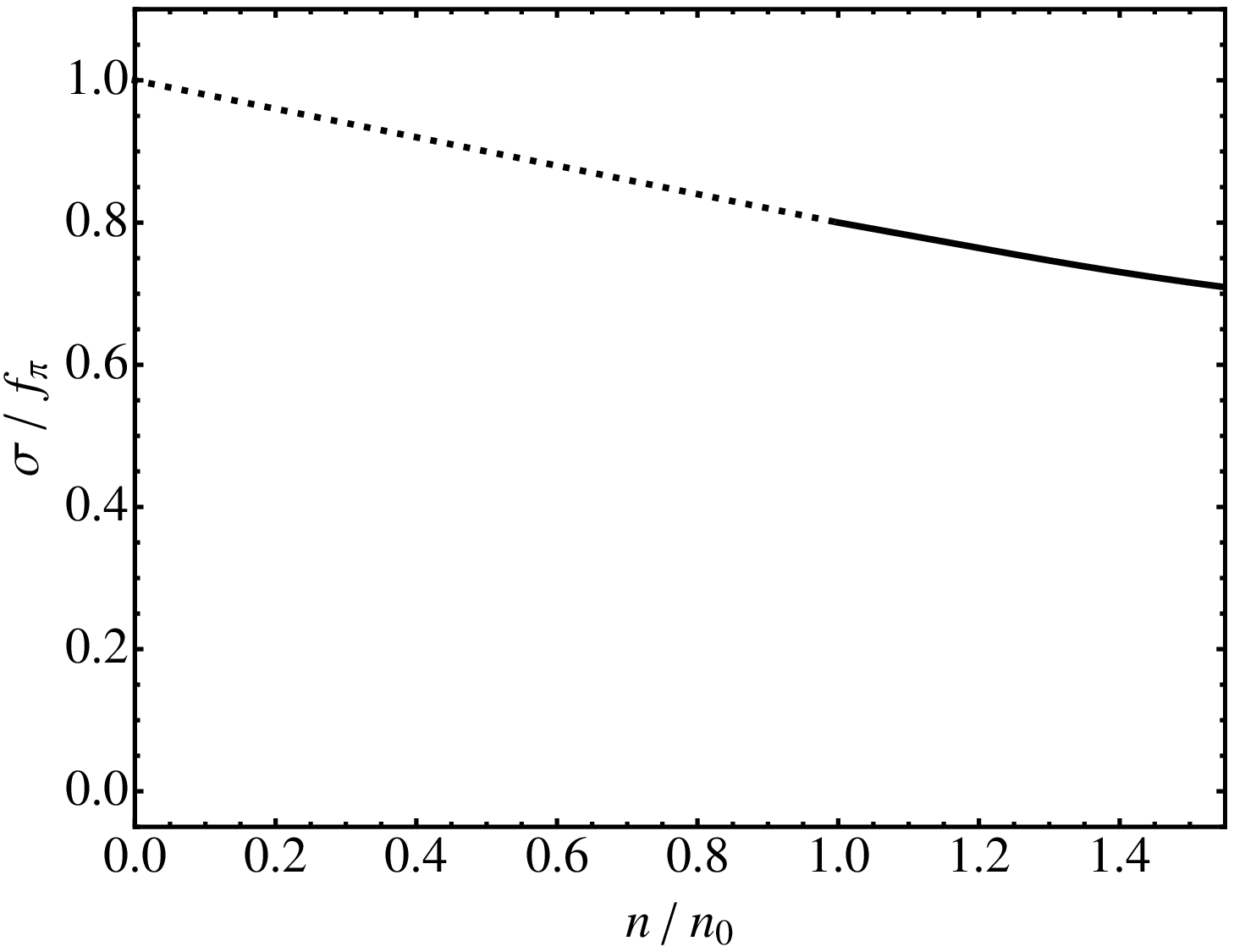}
		\put(80,50){FRG}
	\end{overpic}
	\qquad
	\begin{overpic}[width=0.4\textwidth]{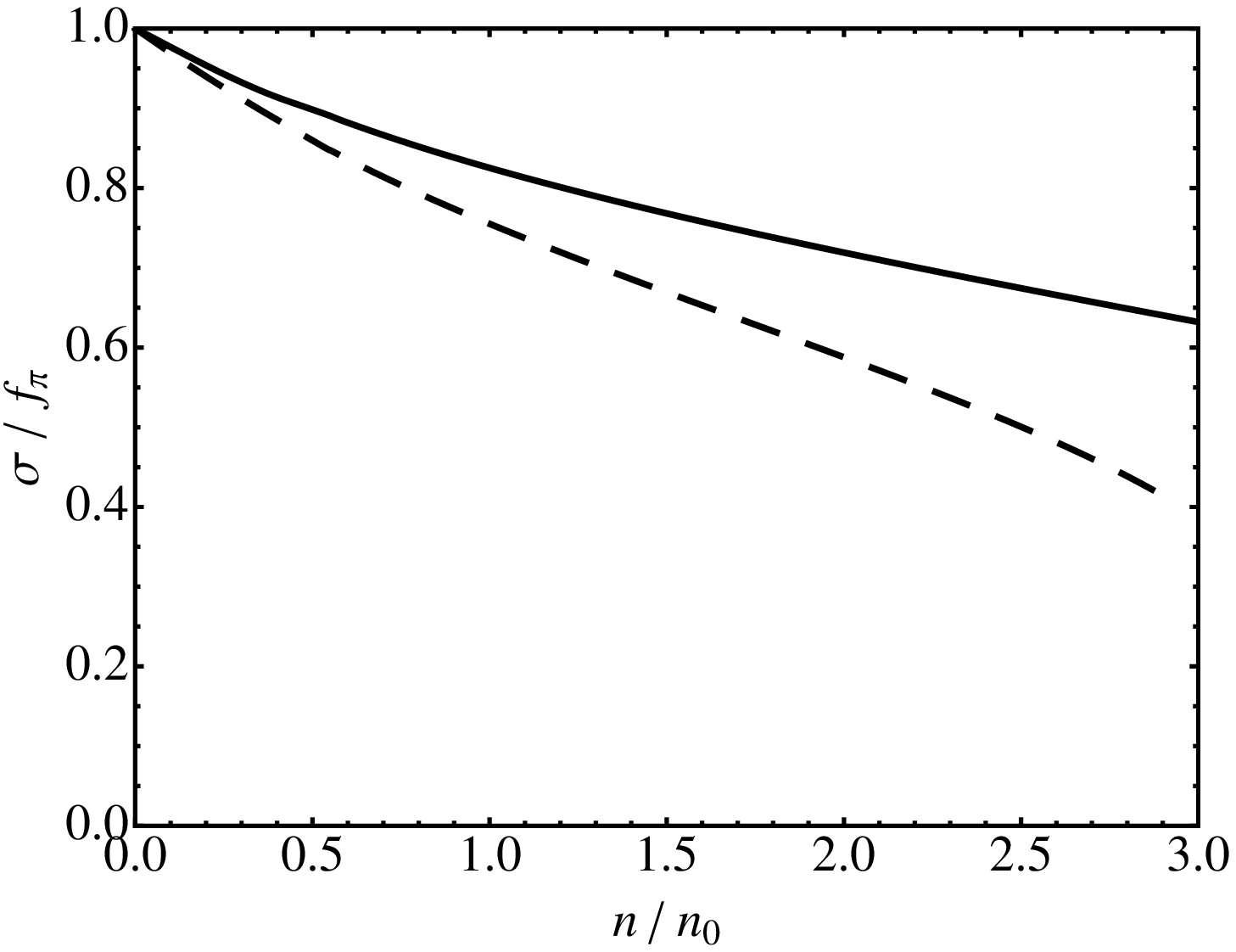}
		\put(70,58){FRG}
		\put(75,47){MF}
	\end{overpic}
	\caption{The chiral order parameter as a function of density (normalized to nuclear saturation density $n_0$). Left: symmetric nuclear matter. Right: pure neutron matter, with FRG-ChNM results (solid line, FRG) and the mean-field results (dashed line, MF).}
	\label{fig:sigma}
\end{figure}

The chiral order parameter is given in this model by the expectation value of the field $\sigma$. On the left-hand side of Fig.\,\ref{fig:sigma}, we show this chiral order parameter normalized to its vacuum value (the pion decay constant $f_\pi$) as a function of density in the case of symmetric nuclear matter. We see that the condensate decreases quite slowly, in agreement with findings from chiral effective field theory \cite{fiorilla2012nuclear}. It is found that the chiral order parameter $\sigma$ stays above sixty percent of its vacuum value $f_\pi$ for temperatures $T\lesssim100\text{\,MeV}$ and baryon chemical potentials up to about $1\text{\,GeV}$. As a consequence, chiral symmetry restoration is moved far away from the well established nuclear physics region around the liquid-gas phase transition. 

When applied to neutron matter, the mean-field approximation of the ChNM model suggests, at first sight, a relatively steep decrease of the chiral order parameter as shown on the right-hand side of Fig.\,\ref{fig:sigma}. In contrast, once fluctuations are included using the functional renormalization group, the chiral condensate is stabilized and chiral symmetry restoration is shifted to much higher densities, typically about seven to eight times nuclear saturation density. Thus our model is capable of dealing with even the inner core of neutron stars in terms of hadronic (rather than quark-gluon) degrees of freedom.

\begin{figure}
	\centering
	\begin{overpic}[width=0.45\textwidth]{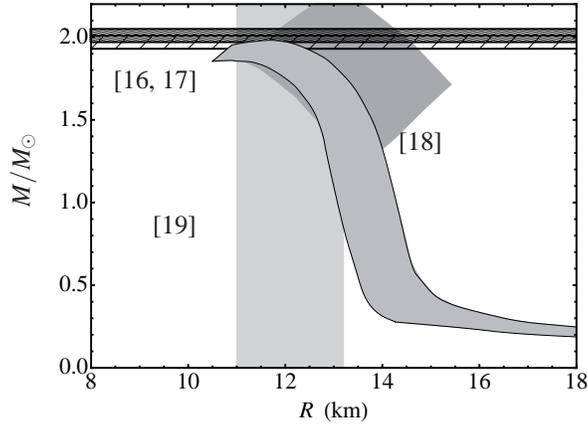}
		\put(15,61){\cite{demorest2010two-solar-mass,antoniadis2013massive}}
		\put(66,50){\cite{truemper2011observations}}
		\put(22,35){\cite{lattimer2014constraints}}
		\put(0,40){\begin{rotate}{90} $M/M_{\odot}$ \end{rotate}}
	\end{overpic}
	\caption{Calculated mass-radius relation for neutron star matter. The band of trajectories in the M-R plane corresponds to a range of values for the nuclear symmetry energy between 29 and 33 MeV. Shown for comparison are the mass-radius constraints from ref.\cite{truemper2011observations}, the radius constraint \cite{lattimer2014constraints}, and the two-solar-mass neutron stars \cite{demorest2010two-solar-mass,antoniadis2013massive}.}
	\label{fig:m_r}
\end{figure}

Given an equation of state as input, the mass-radius relation of neutron stars can be computed using the Tolman--Oppenheimer--Volkoff (TOV) equations. As a test case we start from a simple scenario in which the crust is described by a phenomenological equation of state while the whole interior of the neutron star is modeled by the FRG-ChNM approach, including beta equilibrium to control the admixture of protons. In Fig.\,\ref{fig:m_r}, we show the mass-radius relation obtained from a solution of the TOV equations, allowing for a range of possible values for the symmetry energy. The most massive stars in our calculation are consistent with the observations of two-solar-mass neutron stars \cite{demorest2010two-solar-mass,antoniadis2013massive}, and also with (less stringent) constraints from radius determinations \cite{truemper2011observations,lattimer2014constraints}.
Because of the required stiffness of the equation of state, the density reached in the core of a neutron star with mass $M\simeq 2\,M_\odot$ is not ultrahigh,  just slightly above five times nuclear saturation density. As pointed out, chiral symmetry is still in its spontaneously broken Nambu--Goldstone realization at these densities, so that a necessary condition for the applicability of the present model is fulfilled. Of course, a more detailed investigation requires considering also hyperons and their interactions in dense matter. 

In summary, the FRG-ChNM model is well suited to study the thermodynamics of both symmetric and asymmetric nuclear matter over a broad range of densities and temperatures. In order to draw conclusions about chiral symmetry restoration in baryonic matter, it is of crucial importance to properly include fluctuations beyond the frequently used mean-field approximation.

%%%%%%%%%%%%%%%%%%%%%%%%%%%%%%%%%%%%%%%%%%%%%%%%
%% BACKMATTER
%%%%%%%%%%%%%%%%%%%%%%%%%%%%%%%%%%%%%%%%%%%%%%%%

\begin{theacknowledgments}
  This work is supported in part by BMBF and by the DFG Cluster of Excellence ``Origin and Structure of the Universe.''
\end{theacknowledgments}


\bibliographystyle{aipproc}   % if natbib is available

%\bibliography{sample}
\bibliography{biblio}{}

\begin{thebibliography}{19}
\expandafter\ifx\csname natexlab\endcsname\relax\def\natexlab#1{#1}\fi
\providecommand{\enquote}[1]{``#1''}
\expandafter\ifx\csname url\endcsname\relax
  \def\url#1{\texttt{#1}}\fi
\expandafter\ifx\csname urlprefix\endcsname\relax\def\urlprefix{URL }\fi
\providecommand{\eprint}[2][]{\url{#2}}

\bibitem[Berges et~al.(2000)]{berges2000chiral}
J.~Berges, D.~U. Jungnickel, and C.~Wetterich, \emph{Eur. Phys. J. C}
  \textbf{13}, 323--329 (2000), \eprint{arXiv:hep-ph/9811347}.

\bibitem[Berges et~al.(2003)]{berges2003quark}
J.~Berges, D.~Jungnickel, and C.~Wetterich, \emph{Int. J. Mod. Phys. A}
  \textbf{18}, 3189--3219 (2003), \eprint{arXiv:hep-ph/9811387}.

\bibitem[Floerchinger and Wetterich(2012)]{floerchinger2012chemical}
S.~Floerchinger, and C.~Wetterich, \emph{Nucl. Phys. A} \textbf{890-891},
  11--24 (2012), \eprint{arXiv:1202.1671}.

\bibitem[Kaiser and Weise(2009)]{kaiser2009chiral}
N.~Kaiser, and W.~Weise, \emph{Phys. Lett. B} \textbf{671}, 25--29 (2009),
  \eprint{arXiv:0808.0856}.

\bibitem[Kr\"{u}ger et~al.(2013)]{krueger2013neutron}
T.~Kr\"{u}ger, I.~Tews, K.~Hebeler, and A.~Schwenk, \emph{Phys. Rev. C}
  \textbf{88}, 025802 (2013), \eprint{arXiv:1304.2212}.

\bibitem[Drews et~al.(2013)]{drews2013thermodynamic}
M.~Drews, T.~Hell, B.~Klein, and W.~Weise, \emph{Phys. Rev. D} \textbf{88},
  096011 (2013), \eprint{arXiv:1308.5596}.

\bibitem[Drews and Weise(2014)]{drews2014functional}
M.~Drews, and W.~Weise, \emph{Phys. Lett. B} \textbf{738}, 187--190 (2014),
  \eprint{arXiv:1404.0882}.

\bibitem[Wetterich(1993)]{wetterich1993exact}
C.~Wetterich, \emph{Phys. Lett. B} \textbf{301}, 90--94 (1993).

\bibitem[Drews et~al.(2014)]{drews2014dense}
M.~Drews, T.~Hell, B.~Klein, and W.~Weise, \emph{{EPJ} web conf.} \textbf{66},
  04008 (2014), \eprint{arXiv:1307.6973}.

\bibitem[Akmal et~al.(1998)]{akmal1998equation}
A.~Akmal, V.~R. Pandharipande, and D.~G. Ravenhall, \emph{Phys. Rev. C}
  \textbf{58}, 1804--1828 (1998), \eprint{arXiv:nucl-th/9804027}.

\bibitem[Armani et~al.(2011)]{armani2011recent}
P.~Armani, A.~Y. Illarionov, D.~Lonardoni, F.~Pederiva, S.~Gandolfi, K.~E.
  Schmidt, and S.~Fantoni, \emph{J. Phys.: Conf. Ser.} \textbf{336}, 012014
  (2011), \eprint{arXiv:1110.0993}.

\bibitem[Hell and Weise(2014)]{hell2014dense}
T.~Hell, and W.~Weise, \emph{Phys. Rev. C} \textbf{90}, 045801 (2014),
  \eprint{arXiv:1402.4098}.

\bibitem[Roggero et~al.(2014)]{roggero2014quantum}
A.~Roggero, A.~Mukherjee, and F.~Pederiva, \emph{Phys. Rev. Lett.}
  \textbf{112}, 221103 (2014), \eprint{arXiv:1402.1576}.

\bibitem[Lattimer and Lim(2013)]{lattimer2013constraining}
J.~M. Lattimer, and Y.~Lim, \emph{{ApJ}} \textbf{771}, 51 (2013),
  \eprint{arXiv:1203.4286}.

\bibitem[Fiorilla et~al.(2012)]{fiorilla2012nuclear}
S.~Fiorilla, N.~Kaiser, and W.~Weise, \emph{Phys. Lett. B} \textbf{714},
  251--255 (2012), \eprint{arXiv:1204.4318}.

\bibitem[Demorest et~al.(2010)]{demorest2010two-solar-mass}
P.~B. Demorest, T.~Pennucci, S.~M. Ransom, M.~S.~E. Roberts, and J.~W.~T.
  Hessels, \emph{Nature} \textbf{467}, 1081--1083 (2010),
  \eprint{arXiv:1010.5788}.

\bibitem[{Antoniadis et al.}(2013)]{antoniadis2013massive}
J.~{Antoniadis et al.}, \emph{Science} \textbf{340}, 1233232 (2013),
  \eprint{arXiv:1304.6875}.

\bibitem[Tr\"{u}mper(2011)]{truemper2011observations}
J.~E. Tr\"{u}mper, \emph{Prog. Part. Nucl. Phys.} \textbf{66}, 674--680 (2011).

\bibitem[Lattimer and Steiner(2014)]{lattimer2014constraints}
J.~M. Lattimer, and A.~W. Steiner, \emph{{EPJ} A} \textbf{50}, 40 (2014),
  \eprint{arXiv:1403.1186}.

\end{thebibliography}

\IfFileExists{\jobname.bbl}{}
 {\typeout{}
  \typeout{******************************************}
  \typeout{** Please run "bibtex \jobname" to optain}
  \typeout{** the bibliography and then re-run LaTeX}
  \typeout{** twice to fix the references!}
  \typeout{******************************************}
  \typeout{}
 }

\end{document}